# Discovery of a Bloch point quadrupole constituting hybrid topological strings

Fehmi Sami Yasin*[1], Jan Masell[1,2], Yoshio Takahashi[3], Tetsuya Akashi[3], Norio Baba[4], Kosuke Karube[1], Daisuke Shindo[1], Takahisa Arima[1,5], Yasujiro Taguchi[1], Yoshinori Tokura[1,6,7], Toshiaki Tanigaki[3], and Xiuzhen Yu*[1]

[1] RIKEN Center for Emergent Matter Science (CEMS), Wako, Japan.
[2] Institute of Theoretical Solid State Physics, Karlsruhe Institute of Technology (KIT), Karlsruhe, Germany
[3] Research and Development Group, Hitachi Ltd., Hatoyama, Japan
[4] Research Institute for Science and Technology, Kogakuin University, Hachioji, Japan
[5] Department of Advanced Materials Science, University of Tokyo, Kashiwa, Japan.
[6] Department of Applied Physics, University of Tokyo, Tokyo, Japan.
[7] Tokyo College, University of Tokyo, Tokyo, Japan
* Email: fehmi.yasin@riken.jp; yu_x@riken.jp

**Topological magnetic (anti)skyrmions are robust string-like objects heralded as potential components in next-generation topological spintronics devices due to their manipulability via low-energy stimuli such as magnetic fields, heat, and electric/thermal current. While these two-dimensional (2D) topological objects are widely studied, intrinsically three-dimensional (3D) electron-spin real-space topology remains less explored despite its prevalence in bulk magnets. Here, we capture the 3D structure of antiskyrmions in a single-crystal, precision-doped $(Fe_{0.63}Ni_{0.3}Pd_{0.07})_3P$ lamellae using holographic vector field electron tomography at room temperature and zero field. Our measurements reveal hybrid string-like solitons composed of skyrmions with topological number $W = -1$ on the lamellae's surfaces and an antiskyrmion ($W = +1$) connecting them. High resolution images uncover a Bloch point (BP) quadrupole (four magnetic (anti)monopoles) positioned along the rectangular antiskyrmion's four corners (Bloch lines), which enable the observed lengthwise topological transitions. Furthermore, we calculate and compare the energy densities of hybrid strings with ideal (anti)skyrmion strings using micromagnetic simulations, which suggest that this composite (anti)BP structure stabilizes via the subtle interplay between the magnetostatic interaction and anisotropic Dzyaloshinskii-Moriya interaction. The discovery of these hybrid spin textures enables topological tunability, a tunable topological Hall effect, and the suppression of skyrmion Hall motion, disrupting existing paradigms within spintronics.**

## Main

Magnetic (anti)skyrmions[1–4], which cause the topological Hall effect and exhibit current-driven Hall motion, are highly desirable solitons due to their topological protection and their manipulability via low-energy stimuli including magnetic fields[5–8], heat[8,9], and electric/thermal current[10–16]. One of the origins of these spin textures lies within the competition between the ferromagnetic exchange interaction and the Dzyaloshinskii-Moriya interaction (DMI)[8,17]. The former concerns the force on each spin to align with its neighbors, whereas the latter is an antisymmetric exchange interaction between neighboring spins that favors spin canting away from parallel alignment. The resulting real space textures are now well known in 2D. Bloch-type skyrmions (Fig. 1a-b) are proper screw-like spin propagations with cores pointing up/down embedded in a background of opposite orientation, and exhibit either clockwise or counterclockwise vortex-like in-plane structure. They contrast with Néel-type skyrmions (Fig. 1c-d) composed of cycloidal spin propagations twisting radially inward or outward from their core spins. Skyrmions of both types are characterized by an integer topological number $W = \frac{1}{4\pi} \int \int \boldsymbol{m} \cdot \left(\frac{\partial \boldsymbol{m}}{\partial x} \times \frac{\partial \boldsymbol{m}}{\partial y}\right) dx\, dy = -1$, where $\boldsymbol{m} = \frac{\boldsymbol{M}}{|\boldsymbol{M}|}$ is the normalized magnetization[12]. On the other hand, antiskyrmions (Fig. 1e-f) have an in-plane antivortex magnetic structure, i.e. the nature of their domain walls alternate between Bloch and Néel-type. The four Néel-type regions are typically termed Bloch lines[4,8,18] and the spin texture exhibits $W = +1$.

Recent results have confirmed that in three-dimensions (3D), skyrmions are string-like objects[5,6,19–26]. Furthermore, dipole-dipole energy can stabilize hybrid skyrmion strings[27–31] composed of Néel-skyrmion surface layers linked via a Bloch-type skyrmion (simulated in-plane magnetization, defined as the region where $|m_z| < 1/6$, is shown in Fig. 1g), and predicted to have lower velocities and skyrmion Hall angles than their 2D counterparts[16,28]. The idea for Néel-type domain wall twists near sample surfaces was introduced over fifty years ago in magnetic bubble-hosting uniaxial ferromagnets[32,33]. It can be understood from the demagnetization field $\mu_0 \boldsymbol{H}_{demag}(\boldsymbol{r})$ generated to cancel the magnetic surface and volume charges,

$$\boldsymbol{B}_{demag}(\boldsymbol{r}) = \mu_0 \boldsymbol{H}_{demag}(\boldsymbol{r}) = -\boldsymbol{\nabla}\left(\frac{1}{4\pi}\left[\int \frac{\rho_V(\boldsymbol{r}')}{|\boldsymbol{r}-\boldsymbol{r}'|} dV' + \int \frac{\sigma_s(\boldsymbol{r}')}{|\boldsymbol{r}-\boldsymbol{r}'|} dS'\right]\right), \quad (1)$$

where $\rho_V = -(\boldsymbol{\nabla} \cdot \boldsymbol{M}(\boldsymbol{r}))$ is the reduced volume charge density and $\sigma_s = \boldsymbol{M}(\boldsymbol{r}) \cdot \boldsymbol{n}$ is the surface charge density, where $\boldsymbol{n}$ is the surface normal. The in-plane component of $\boldsymbol{B}_{demag}(\boldsymbol{r})$ perpendicular to domain walls increases substantially near the surface, torquing the magnetic bubble domain wall magnetization field to be entirely Néel-like for some depth below the surface[34].

In $D_{2d}$ and $S_4$ crystal symmetry magnets, DMI is anisotropic, leading to the formation of antiskyrmions, skyrmions and nontopological bubbles[4,8,17,35]. Moreover, dipole-dipole interactions combine with anisotropic DMI to form characteristic square-shaped antiskyrmions[8]. In $(Fe_{0.63}Ni_{0.3}Pd_{0.07})_3P$ (FNPP), (anti)skyrmions were observed to have a vortex-like magnetic field configuration within their cores[17], suggesting a more complex 3D structure than the 2D model assumes. Micromagnetic simulations including demagnetization suggest the Bloch regions of an antiskyrmion's domain walls twist Néel-type near the surfaces (Fig. 1h), deforming the spin texture: the Bloch-type regions widen while the Bloch line magnetizations (four corners of Fig. 1h) widen on one surface and narrow on the other.

Recent work has also revealed controlled transformations between topological skyrmions and antiskyrmions via nontopological bubbles[8]. Such transformations are enabled by micromagnetic structures called (anti)Bloch points (BPs), which are singularities in the magnetization with (anti)hedgehog structure (Fig. 2b-e). These magnetic monopoles have been predicted to play a role in magnetization dynamics including its reversal in soft ferromagnets[36], pinning to the atomic lattice[37], domain wall dynamics[38], skyrmion lattice annihilation[39] and vortex core reversal[40]. BPs have also been found to reduce the total energy of a thin film sample when present within a singular Bloch line that switches its orientation along the sample[41,42]. While (anti)BPs have been observed in real space using both X-rays[19,43,44] and electrons[6], the detailed 3D structure has yet to be measured in high resolution. Similarly, hybrid topological spin textures have yet to be confirmed experimentally via real space 3D imaging.

In this work, we use off-axis electron holography configured for vector field electron tomography (VFET)[20,45–47] to image a hybrid antiskyrmion string (Fig. 2a) in an FNPP nanoplank at room temperature and zero field. These 3D measurements unveil four (anti)BPs located along the antiskyrmion's Bloch lines and arranged in a quadrupole, enabling a topological transformation from skyrmion at the surfaces to antiskyrmion in the bulk. Furthermore, energy densities calculations of these hybrid strings suggest that the dipolar and DMI energy terms are leading contributors in stabilizing topologically complex hybrid strings over their ideal counterparts.

**Stabilization of hybrid strings in FNPP**
Hybrid strings naturally stabilize as the lowest energy density nontrivial topological spin texture when relaxing an antiskyrmion state using micromagnetic simulations with FNPP parameters[17,35]. As shown in Fig. 2a, the in-plane magnetization depicts a domain wall twisting from Néel-type skyrmions on the surfaces to antiskyrmion in the center. To enable this transition, (anti-)BPs form near the antiskyrmion's four corners: two BPs near the bottom surface transforming the skyrmion into an antiskyrmion, Figs. 2(b, d), and two antiBPs near the top surface transforming the antiskyrmion into a skyrmion, Figs. 2(c, e). This topological transition is also illustrated in the magnetic field $\boldsymbol{B}(\boldsymbol{r}) = \mu_0 \boldsymbol{M}(\boldsymbol{r}) + \boldsymbol{B}_{\text{demag}}(\boldsymbol{r})$ slices shown in Fig. 2f-j ($xy$-plane view) and Fig. 2k-o ($xz$-plane view). Here, the skyrmions orient radially inward on top (Fig. 2f) and outward on bottom (Fig. 2j), minimizing the magnetostatic (dipole-dipole) energy by closing the magnetic field lines running along the spin texture's core and periphery.

The hybrid string's 3D magnetic field is shown in Fig. 2p, indicating another feature of note: the shrinking of the core near the surfaces and corresponding expansion of the domain wall's width consistent with the combination of weak uniaxial anisotropy and strong dipole-dipole interaction characteristic of FNPP. The positions of the four (anti)BPs are indicated by dashed black crosses in the right-hand-side (RHS) panel of Extended Data Fig. 1a, which shows the center slice of $\boldsymbol{B}(\boldsymbol{r})$. Extended Data Figs. 1b-e plot $xz$- and $yz$-slices of $\boldsymbol{B}(\boldsymbol{r})$ intersecting the (anti)BP positions side-by-side with the magnitude of $\boldsymbol{B}(\boldsymbol{r})$ versus $z$ along the Bloch lines which host the (anti)BPs. Notably, $|\boldsymbol{B}(\boldsymbol{r})|$ decreases to a minimum at the (anti)BP position. This is because $\boldsymbol{M}(\boldsymbol{r})$ has a non-zero divergence $\nabla \cdot \boldsymbol{M}(\boldsymbol{r})$ at the (anti)BPs, as shown in Extended Data Fig. 2a. $\nabla \cdot \boldsymbol{M}(\boldsymbol{r}) > 0$ for the BPs and $\nabla \cdot \boldsymbol{M}(\boldsymbol{r}) < 0$ for the antiBPs, indicating that the magnetization field is source- and sink-like at those points, respectively. To illustrate this, we performed a smooth transformation on the BPs shown in the top panels of Extended Data Fig. 2(b, d) and antiBPs in the top panels of Extended Data Fig. 2(c, e) while conserving the topological number, resulting in the hedgehogs and antihedgehogs shown in the bottom panels. The nonzero $\nabla \cdot \boldsymbol{M}(\boldsymbol{r})$ at the (anti)BP quadrupole generates a nonzero $\nabla \cdot \boldsymbol{B}_{demag}(\boldsymbol{r})$, resulting in $\nabla \cdot \boldsymbol{B}(\boldsymbol{r}) = 0$ (Extended Data Fig. 2f) and decrease in $|\boldsymbol{B}(\boldsymbol{r})|$.

The anisotropy ($E_{ani}/V$), dipolar ($E_{dd}/V$), exchange ($E_{ex}/V$), DMI ($E_{dmi}/V$), and total ($E_{tot}/V$) volume energy densities are plotted as functions of $z$ for hybrid, antiskyrmion, and skyrmion strings in Extended Data Fig. 3a-c, respectively. The gray-shaded thickness region in Extended Data Fig. 3a indicates $W = +1$, while the neighboring regions have $W = -1$. The (anti)BPs are located where W changes sharply from skyrmion-like ($W = -1$) to antiskyrmion-like ($W = +1$) on the top surface (antiBPs) and vice versa on the bottom surface (BPs). At these singular points, $E_{tot}/V$ rises sharply due to the exchange interaction but decreases on the surfaces when compared to $E_{tot}/V$ of an ideal antiskyrmion string due to the skyrmion-like surface spin textures that satisfy the flux loop closure desired by dipolar interactions. Note that ideal antiskyrmion strings never fully optimize this interaction as their in-plane antivortex character always comprises regions with moments pointing inward and outward. A comparison of the energy densities of hybrid strings (solid green), skyrmion strings (dashed blue) and antiskyrmion strings (dotted orange) are plotted as a function of $z$ in Extended Data Fig. 3d. On the other hand, anisotropic DMI favors a bulk antiskyrmion, and so $E_{tot}/V$ of the hybrid string is lower than the ideal skyrmion in this region.

$E_{tot}/V$ of hybrid strings (solid green) is lower than skyrmion (dashed blue) and antiskyrmion strings (dotted orange), as well as the polarized state (solid black) at low magnetic field ($B_{ext}$) values including zero field, as shown in Fig. 2q. Notably, $E_{tot}/V$ of ideal antiskyrmion strings remains higher than that of hybrid strings for all $B_{ext}$ values. As we will discuss later, the introduction of $B_{ext}$ changes the $z$-location of the (anti)BPs, resulting in a tunable thickness-averaged topological number $\overline{W}$. We plot $\overline{W}(B_{ext})$ in Extended Data Fig. 3e for the polarized/helical state (solid black line), ideal skyrmion string (dashed blue line), ideal antiskyrmion string (dotted orange line), and hybrid string (solid green line). Extended Data Fig. 4 shows a comparison of the in-plane magnetic induction between the skyrmion (Extended Data Fig. 4a-c), antiskyrmion (Extended Data Fig. 4d-f), and hybrid string (Extended Data Fig. 4g-i) at the top (Extended Data Fig. 4(a, d, g)), middle (Extended Data Fig. 4(b, e, h)), and bottom (Extended Data Fig. 4(c, f, i)) of the spin textures. The decrease in energy from both the skyrmion capping layers and decreased bulk shape deformation compensates for the energy increase at the (anti)BPs, stabilizing the hybrid string.

**Imaging magnetic monopoles**
We prepared a metastable hybrid string within an FNPP square-tipped nanoplank (Extended Data Fig. 5a-b, see methods and supplementary text for details). We used off-axis electron holography VFET (experimental geometry outlined in Extended Data Fig. 6) to acquire holograms at tilt angles along two tilt axes over a range of 360° in 5° steps. We followed the procedure described in Table S1 to extract the magnetic phase along every tilt angle at which the electron beam could penetrate through the sample without a significant loss of beam coherence due to inelastic scattering events, resulting in a 130° tilt angle range along both tilt axes We reconstructed the 3D magnetic field using the recently developed quantisation units reconstruction technique (QURT). QURT does not require a priori knowledge of the 3D structure and provides higher resolution and smaller z-elongation caused by a missing wedge of tilt angles when compared to other reconstruction algorithms [48,49].

The resulting reconstruction of $\boldsymbol{B}(\boldsymbol{r})$ in the region of the hybrid string is presented in Fig. 3. As predicted in the previously described simulations, the transition from skyrmion on the top surface to antiskyrmion in the bulk to skyrmion on the bottom surface is clearly seen in the $xy$-plane slices shown in Figs. 3a-c. Furthermore, this topological transformation occurs at distinct $z$-slices containing (anti)BPs, two of which are circled in Figs. 3d-e. The $xz$- and $yz$-planes that intersect at the Bloch line along which the (anti)BP circled in Fig. 3d is located are plotted using 3D glyphs in the RHS panel. The magnitude of $\boldsymbol{B}(\boldsymbol{r})$ along the Bloch line (Fig. 3f) decreases to nearly zero at the anti-BP, in agreement with the previously described simulations. The $xz$- and $yz$-planes of $\boldsymbol{B}(\boldsymbol{r})$ that intersect at all four Bloch lines are plotted along with $|\boldsymbol{B}(\boldsymbol{r})|$ as a function of $z$ in Extended Data Fig. 7. The locations of the other (anti)BPs (circled in black) may also be seen from the vortex-like whirls on either side of the downward oriented core spins shown in Fig. 3g-h, which are slices selected to intersect two (anti)BPs simultaneously along the hybrid string's diagonals indicated in Fig. 3b. The length of the arrows/glyphs in Fig. 3a-e corresponds to $|\boldsymbol{B}(\boldsymbol{r})|$, whereas $\boldsymbol{B}(\boldsymbol{r})$ is normalized in Fig. 3g-h to increase overall visibility. Curiously, $\boldsymbol{B}(\boldsymbol{r})$ does not whirl in a vortex-like texture within the displayed plane around the (anti)BP on the left-hand-side (LHS) of Fig. 3h, but instead rotates within the $xy$-plane down the z-direction.

In addition to a 3D plot of normalized $\boldsymbol{B}(\boldsymbol{r})$ in Fig. 4, we plot a slice of the magnetic field dissecting the rectangular hybrid string's long axis in the upper panel using glyphs superimposed on a color plot of the core spins defined as the region in which the $z$ component of the measured magnetic field $B_z(\boldsymbol{r}) < -0.2$. The thin plate's surfaces are indicated on the top and bottom of the slice, and the domain walls are outlined with dashed blue and red lines on the LHS and RHS, respectively. We plot the core area per slice in the upper RHS panel of Fig. 4. The hybrid string's core area shrinks near the surfaces, corresponding to an increase in the domain walls' width, to minimize dipolar energy. The core area of the antiskyrmion portion of the spin texture (the z-center region) is the largest region, with the core's lateral contraction occurring in the skyrmion portions of the spin texture, suggesting that skyrmions are quite malleable, consistent with recent studies on skyrmion deformation[15,27].

**Complex topology of hybrid strings**
A $z$-dependent topological charge leads to various consequences for a hybrid string's stability and dynamics. First, the existence of (anti)BPs begs the question whether their z-location can be controlled, and our micromagnetic simulation results suggest they can indeed be moved via $B_{ext}$, and ultimately eject from the thin plate for large $B_{ext}$. More interestingly, this implies that $\overline{W}$ may vary in a controlled way. Such a topological tuning would significantly alter the hybrid string's electric or thermal current-induced dynamics due to the magnus force, the force perpendicular to the driving current direction ($F \parallel x$), which causes the so-called skyrmion Hall motion. For example, the magnus force in 2D models results in a velocity component perpendicular to the driving force, $v_y \propto W$. If we extend this into 3D and assume that the spin texture is rigid during motion, we can expect $v_y \propto \int_0^t W(z)\,dz$, where $W(z)$ is the $z$-dependent topological charge and $t$ is the sample thickness. We anticipate a comparable impact on transport measurements, as the topological Hall effect (THE) also varies with W. In other words, hybrid antiskyrmion-hosting materials with a small characteristic domain width and large saturation magnetization may exhibit a tunable THE through changes in thickness, $B_{ext}$, or both.

The ability to tune a spin texture's transverse motion or a device's THE signal by simply changing $B_{ext}$ or sample thickness would have wide reaching impacts in spintronics and transport, respectively. Moreover, this Hall motion and THE signal can be eliminated altogether with the right combination of sample thickness and $B_{ext}$, as suggested by Extended Data Fig. 3e. Notably, $\overline{W} \approx -0.11$ for a hybrid string at zero field, which may be interpreted as the skyrmion regions occupying 60% of the spin texture's length, with the other 40% being an antiskyrmion. As $B_{ext}$ increases, the spin texture becomes more skyrmion-like as the (anti)BPs penetrate deeper into the bulk of the sample until some threshold field $B_{ext} \approx 320$ mT, above which the trend reverses. This is also reflected in the energy density of each spin texture plotted in Fig. 2q, where the antiskyrmion string's energy density decreases at a higher rate than both the skyrmion string and hybrid string for large, increasing $B_{ext}$. We therefore propose an experiment investigating the current-driven motion of hybrid strings and predict that for thin ($t \leq 140$ nm) plate devices, hybrid strings exhibit skyrmion-like Hall motion for a wide range of applied magnetic fields, and that the motion may be tuned towards antiskyrmion-like using thicker devices or materials with a stronger anisotropic DMI.

In this work, we used holographic VFET to reveal the 3D structure of a (anti)BP quadrupole constituting a zero-field hybrid string. We performed this imaging months after stabilizing the spin texture in the FNPP nanoplank, demonstrating its long lifetime and robust metastability. Another intriguing research direction is what becomes of the (anti)BPs upon application of an electric/thermal current. These topological transition points may act as pinning defects that must be overcome, or perhaps they eject from the system upon application of a driving current. Either way, these points are essential for the construction of hybrid strings and require further study.

**Methods**
**Sample preparation.**
Single-crystalline bulk samples of $(Fe_{0.63}Ni_{0.3}Pd_{0.07})_3P$ were synthesized by a self-flux method from pure Fe, Ni and Pd metals and red phosphorous sealed in an evacuated quartz tube. The target phase of tetragonal $M_3P$ was isolated from the ingot. Phase purity of the $M_3P$ structure was confirmed by powder X-ray diffraction with Cu Kα radiation[17]. Crystal orientations were checked by an X-ray Laue diffraction method. Chemical compositions were examined by a scanning electron microscope equipped with an energy dispersive X-ray analyzer[17].

**Nanoplank fabrication.**

We performed focused ion beam (FIB) lift out using the Thermo Scientific Helios 5 UX DualBeam. After lifting out a 3 µm thin plate onto a TEM Cu half-moon-shaped mesh, we thinned the plate to approximately 500 nm, then rotated the sample 90° to cut the desired plank shape with a 1 µm × 1 µm square end. We chose a square-tipped sample geometry to help stabilize the target square-shaped antiskyrmions[50]. We transferred the sample to a 3D needle via FIB liftout for mounting in a TEM sample holder capable of 360° tilting. We rotated the sample −90° to its original position to thin the plate to the desired thickness. After the final fabrication, we used a 5 keV ion beam to polish the two flat surfaces to remove any Ga implantation that may have occurred. After polishing, we measured the final thickness using the scanning electron microscope (SEM) to be $t \approx 140$ nm.

**Antiskyrmion creation procedure.**
We transferred the sample to a 3D needle via FIB liftout for mounting in a TEM sample holder capable of 360° tilting. We applied a magnetic field $B_{ext} \approx 115$ mT along the [001] crystal axis and field cooled from 410 K to 295 K to create a metastable mixed spin texture lattice composed of mostly antiskyrmions, but also including some skyrmions and non-topological bubbles. We decreased the field to zero, after which several spin textures remained metastable, including the single antiskyrmion embedded in a helical background shown in Extended Data Fig. 5c. We then used FIB to lift out the nanoplank and attach it to a Hitachi TEM tomography holder needle capable of 360° tilting along two perpendicular axes. The antiskyrmion remained metastable in the nanoplank for months after transferring to the needle and before performing 3D imaging, highlighting the long lifetime of the spin texture.

**Holographic vector field electron tomography.**
We performed holographic vector field electron tomography in a 1 MV Hitachi holography electron microscope[51] and configured a double-biprism experimental setup[52] to achieve high precision phase measurements. In off-axis electron holography, a plane wave of electrons transmits through the sample while a portion of the incident wave travels through vacuum as a reference wave. The beam is then split by a Möllenstedt electrostatic biprism, biased such that the sample and reference waves are sufficiently phase shifted to form a hologram at the camera, from which the sample phase may be reconstructed.

**Micromagnetic simulations.**
For the theoretical analysis we consider the same micromagnetic (continuum) model and parameters as determined at $T = 300$ K in Ref. [17]. The energy functional $E[\boldsymbol{m}] = \int \mathcal{E}[\boldsymbol{m}]\, dV$ comprises the magnetic stiffness $A_{ex} = 8.1$ pJ m$^{-1}$, an $S_4$-symmetric Dzyaloshinskii-Moriya interaction (DMI) $D = 0.2$ mJ m$^{-2}$ which favors right-handed Bloch-type domain walls or helices in the x-direction and left-handed ones in the y-direction, uniaxial anisotropy $K_u = 31$ kJ m$^{-3}$, the Zeeman interaction with an external field $B_{\text{ext}}$, and the demagnetizing field $\boldsymbol{B}_{demag}(\boldsymbol{r})$ due to the saturation magnetization $M_s = 417$ kA m$^{-1}$. Note that the real space data presented in this study depicts a 90-degree rotated sample to match the visualization of the experimental data. Explicitly, the energy density reads:

$$E[\boldsymbol{m}] = A\, (\nabla \boldsymbol{m})^2 + D\, [\boldsymbol{m} \cdot (\hat{x} \times \partial_x \boldsymbol{m}) - \boldsymbol{m} \cdot (\hat{y} \times \partial_y \boldsymbol{m})] - K_u\, (\boldsymbol{m} \cdot \hat{z})^2 - \boldsymbol{m} \cdot B_{\text{ext}} - \frac{M_s}{2} \boldsymbol{m} \cdot \boldsymbol{B}_{\text{demag}}$$

We use our modified version of MuMax3[53] to relax initial guesses for the magnetization texture and thereby find local energy minima of the energy $E[\mathbf{m}]$. The modifications which we applied to MuMax3 include the specific form of the DMI and an upgrade for the discretization of derivatives, which in our code are approximated by higher order finite difference schemes which involve more neighbouring sites, as previously used in Refs. ([6,17,35]). This results in slightly higher runtimes but much improved numerical accuracy (except very close to the Bloch points (BP) where the numerical error does not improve with this approach).

To minimize the energy density of the various magnetic phases, i.e., (i) helices, (ii) skyrmions, (iii) antiskyrmions, and (iv) hybrid (anti)skyrmions, we initialize these textures on a numerical lattice with $256 \times 256 \times 128$ lattice sites which yields very high numerical accuracy. The lattice constant in the $z$-direction is chosen such that the thickness is 140 nm to match the experimentally measured specimen thickness. The boundary conditions are periodic in the $x$- and $y$-direction and von-Neumann (open) in the $z$-direction, using MuMax3's built-in function *setPBC(16,16,0)*. For a given value of the magnetic field $B_{\text{ext}} = 0, 50, 100, \dots, 400$ mT the initial textures are optimized in independent simulations using *relax()* for different system sizes $L_x = L_y = 250, 260, 270, \dots, 600$ nm. The energy density $E^{\text{dens}} \equiv E/L_x L_y L_z = E/V$ (energy per volume) for every system size is then obtained by subtracting the energy of the $z$-polarized state from the energy of the relaxed state and dividing by the respective system size. The states with lowest energy density are considered the thermodynamic ground states in this effective $T = 0$ model and their energy densities are compared in Fig. S3D. For the calculation of the topological number, we used the lattice version by Berg and Lüscher[54].

We note that the singular BP configurations, by definition, violate the assumption of smoothly varying magnetization underlying the continuum approximation in the micromagnetic model. However, the physical error is probably small as the dominant energy contributions come from the magnetization around the BP which is distorted on length scales of the micromagnetic model. Moreover, we checked that the numerical error is small by comparing simulation results of ideal BPs to the analytical solution for the case of only magnetic stiffness. Another possible issue arising from BPs on a numerical lattice is artificial pinning. This is indeed problematic as we must initialize the hybrid textures with BPs which during the relaxation procedure are subject to pinning forces on the lattice. We checked that this effect is small by initializing hybrid antiskyrmion strings with various values for the thickness of the skyrmion surface caps. Different initial cap thicknesses did not alter the final cap thickness very much. Moreover, we found that these surface caps also naturally appear in many simulations when the initial state is a homogeneous antiskyrmion string, underlining that the formation of the cap states is a robust mechanism. However, we also realized that the fine numerical mesh chosen for the final plots in this study further increased the mobility of the BPs on the numerical lattice. In previous tests with two times lower numerical resolution and significantly faster runtimes we found that hybrid antiskyrmion strings tend to remain in the $S_4$-symmetric state in which they were created, i.e., the antiskyrmions stay square-shaped. Upon decreasing the lattice constants, we found that the BPs become sufficiently mobile to allow for spontaneous symmetry breaking towards rectangular antiskyrmions due to pair-wise attraction of BPs on distinct interfaces.

**Acknowledgments:**
We are very grateful to Tomoko Kikitsu (Materials Characterization Support Team in the RIKEN Center for Emergent Matter Science) for technical support on the TEM (JEM-2100F), as well as Ilya Belopolski, Max Hirschberger, Max Birch, and Naoto Nagaosa for helpful discussions. Y.Tokura acknowledges the support of Japan Science and Technology Agency (JST) CREST program (Grant Number JPMJCR1874). X.Z.Y. acknowledges the support of Grants-In-Aid for Scientific Research (A) (Grant No. 19H00660) from JSPS and JST-CREST program (Grant No. JPMJCR20T1). J.M. was supported by the Alexander von Humboldt Foundation as a Feodor Lynen Return Fellow.

**Author contributions:**

FSY, XZY, Y. Tokura, and Y. Taguchi conceptualized the project. KK and Y. Taguchi grew the bulk single crystal. FSY performed all device fabrication. FSY, TT, Y. Takahashi, and T. Akashi and XZY designed the experiment. T. Akashi performed the real space imaging. Y. Takahashi processed the raw images. NB perfomed the 3D reconstruction. FSY processed the final 3D images and made visualizations for the figures, with input from JM and T. Arima. JM performed all micromagnetic simulations and energy calculations, with input from FSY. TT, XZY, Y. Taguchi, DS, T. Arima, and Y. Tokura oversaw the project. FSY and JM wrote the original manuscript draft. All authors discussed the results and commented on the manuscript.

**Competing interests:** Authors declare that they have no competing interests.

**Data availability:** All data are available in the main text or the supplementary materials. All other data are available from the corresponding authors upon reasonable request.

**Code availability**
The code used to generate the micromagnetic simulations should be reproducible from the descriptions provided in the 'Micromagnetic simulations' section. Otherwise, the code is available from the corresponding author upon reasonable request.

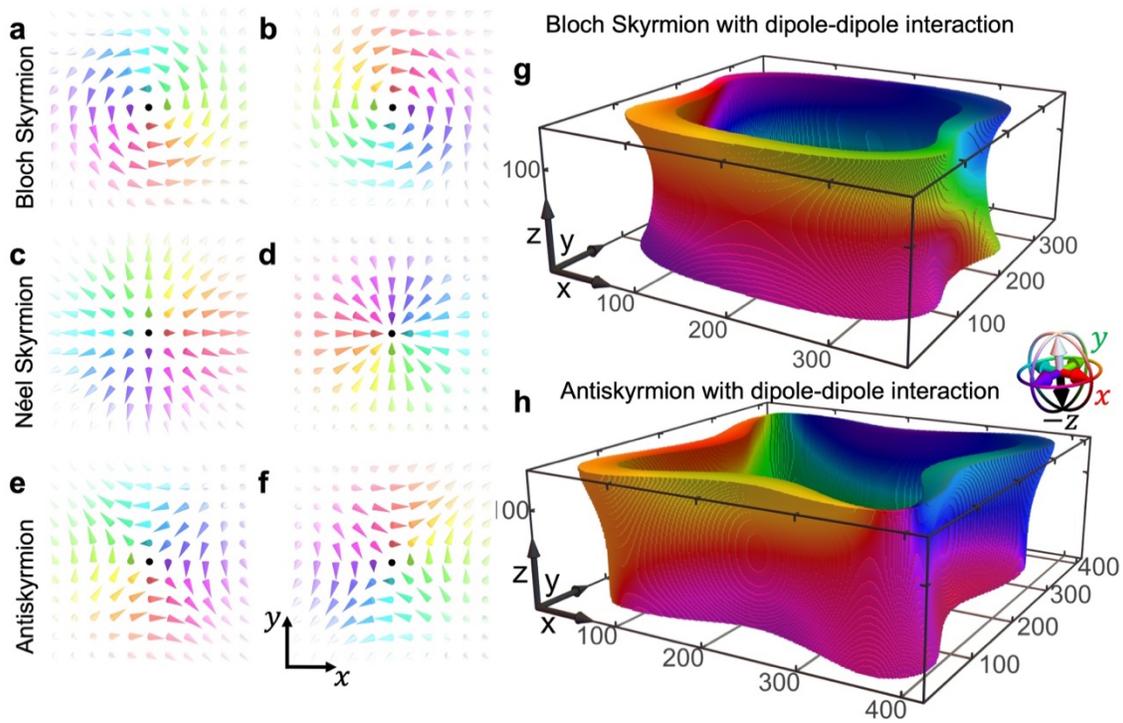

**Figure 1: Real space structure of skyrmions and antiskyrmions.** (**a-b**) Bloch-type skyrmions with topological number $W = -1$ and helicity (**a**) $\gamma = \pi/2$ and (**b**) $\gamma = -\pi/2$. (**c-d**) Néel-type skyrmions with topological charge $W = -1$ and helicity (**c**) $\gamma = 0$ and (**d**) $\gamma = \pi$. (**e-f**) Antiskyrmions with topological charge $W = +1$ and helicity (**e**) $\gamma = -\pi/2$ and (**f**) $\gamma = \pi/2$. (**g-h**) In-plane magnetization ($|m_z| < 1/6$) of ideal (**g**) skyrmion and (**h**) antiskyrmion strings stabilized using micromagnetic simulations including dipole-dipole interactions. The axis labels are in units of length [nm].

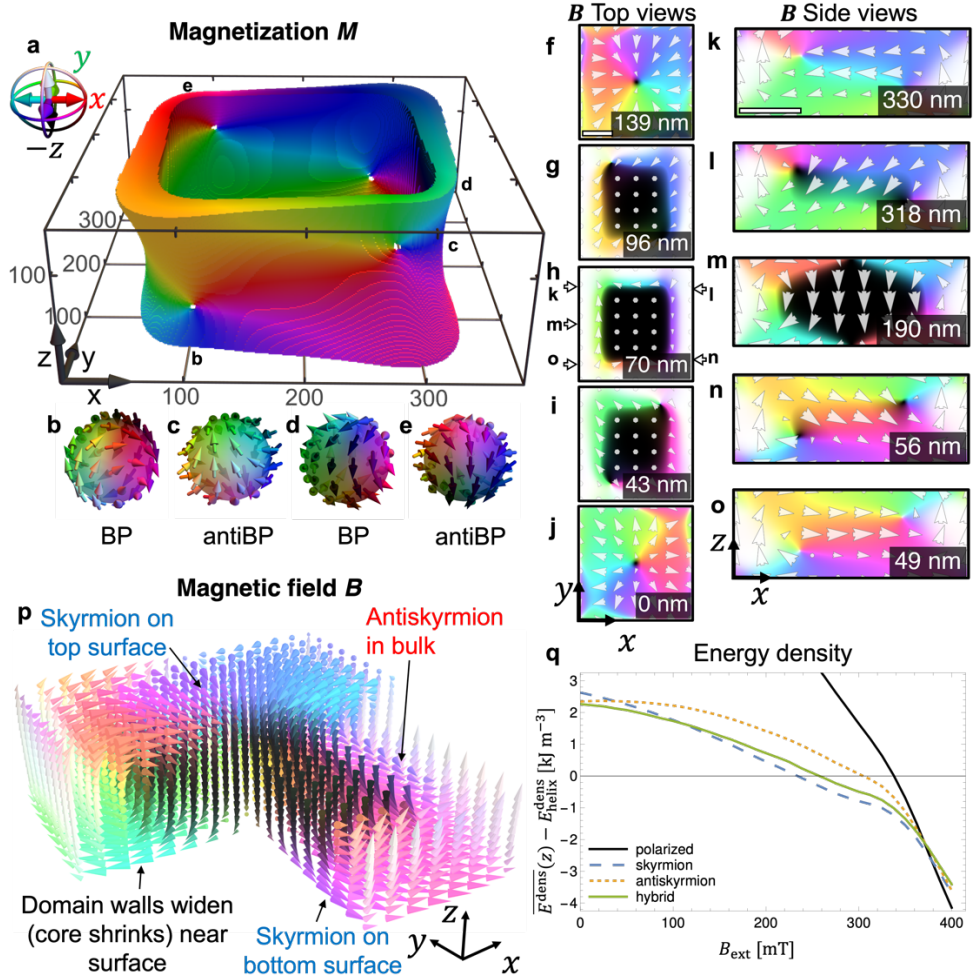

**Figure 2: Hybrid antiskyrmion strings and their three-dimensional (3D) features.** (**a**) Isosurface depicting the in-plane magnetization ($|m_z| < 1/6$) of a micromagnetically simulated hybrid antiskyrmion string with thickness 140 nm and optimized energy density. The axis labels are in units of length [nm]. (**b-e**) BPs and antiBPs indicated in (**a**) magnified with both color and glyph plots. These points indicate the depth from the surface at which the hybrid antiskyrmion string topologically transitions from a skyrmion at the surface to an antiskyrmion in the bulk. (**f-j**) $xy$-slices of the hybrid antiskyrmion string's magnetic field $B$ at (**f**) $t = 139$ nm, (**g**) 96 nm, (**h**) 70 nm, (**i**) 43 nm and (**j**) 0 nm. (**k-o**) $xz$-slices of the hybrid antiskyrmion string's magnetic field at (**k**) $y = 330$ nm, (**l**) 318 nm, (**m**) 190 nm, (**n**) 56 nm and (**o**) 49 nm. (**p**) 3D glyph plot of $B$ with selected regions omitted to show cross-sections of the bulk of the spin texture simultaneously with the surface. (**q**) Energy density as function of magnetic field $B_{ext}$, respectively. The hybrid solution is the first excited below a critical field threshold, outperforming ideal (anti)skyrmions. Scalebars in **f**, **k** are 100 nm.

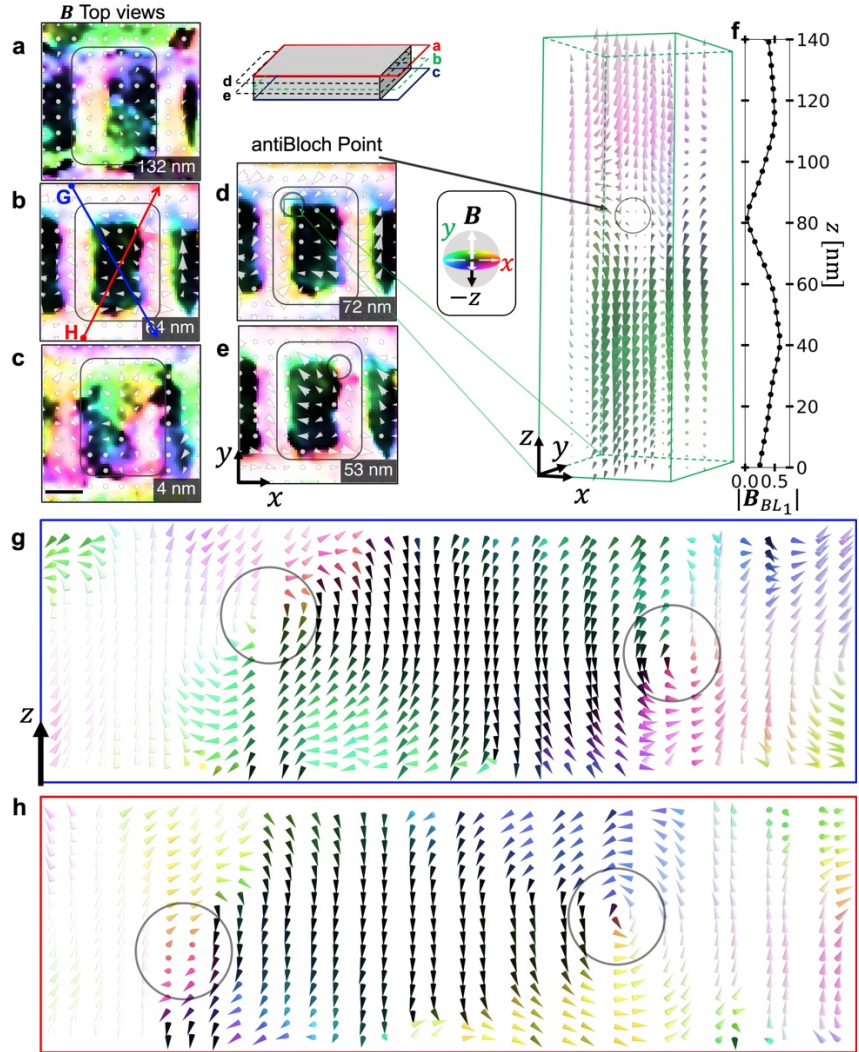

**Figure 3: Vector field tomography reconstruction of a hybrid antiskyrmion string in (Fe$_{0.63}$Ni$_{0.3}$Pd$_{0.07}$)$_3$P.** (**a-e**) $xy$-slices of the experimentally measured magnetic field at (**a**) $t = 132$ nm, (**b**) 64 nm, (**c**) 4 nm showing the surface skyrmions (**a, c**) and bulk antiskyrmion (**b**), respectively, as well as at (**d**) $t = 79$ nm and (**e**) 53 nm which indicate the spatial positions of the (anti)BPs (circled with a solid black line) where the spin texture transforms topologically from skyrmion to antiskyrmion and vice-versa. The spin texture is outlined by a solid dark grey rectangle in **a-e**. The right-hand-side inset plots the magnetic field in the $xz$- and $yz$-planes intersecting the antiBP indicated by a solid green square in **d**. (**f**) Plot of the magnitude of the magnetic field measured along the Bloch line ($|B_{BL_1}|$) as a function of $z$. (**g-h**) Glyph plots of the normalized magnetic field along the Bloch lines within the sliced planes (intersecting the $xy$-plane) indicated by blue (**g**) and red (**h**) rectangles in **b**. The (anti)BP locations are circled in solid black. Scalebar in **c** is 100 nm.

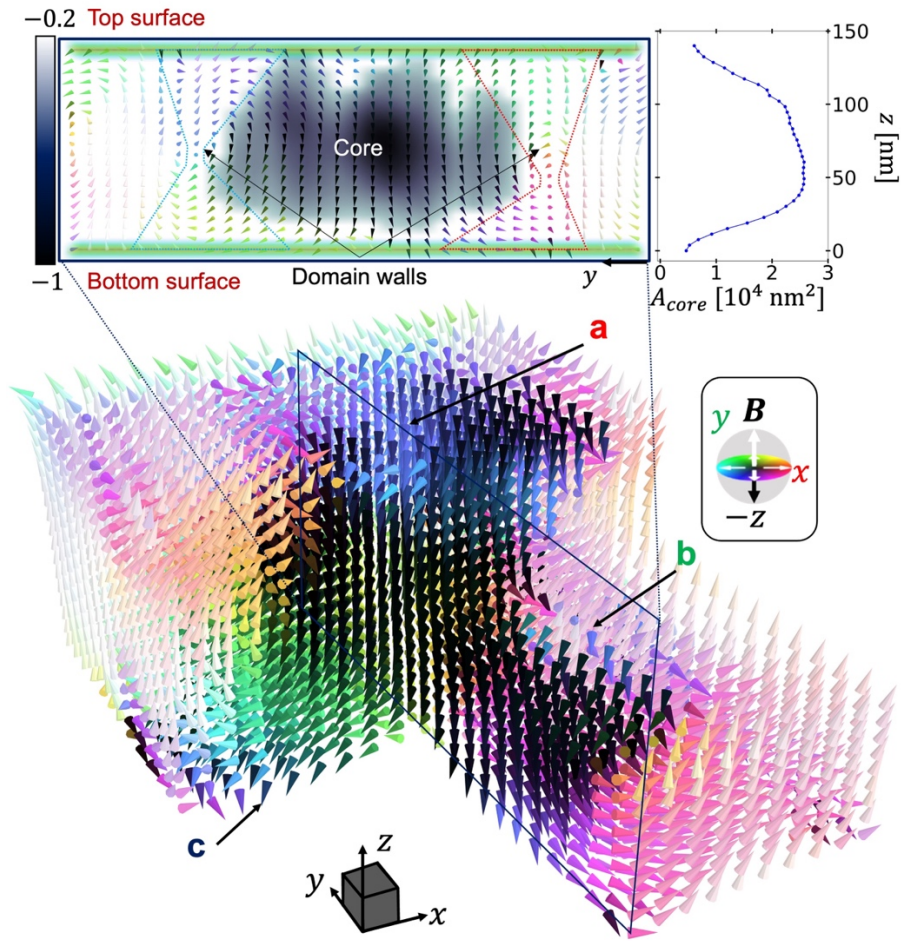

**Figure 4: Experimentally measured three-dimensional (3D) spin structure of a hybrid antiskyrmion string.** Lower: 3D glyphs of the normalized magnetic field binned to 33 voxels per dimension and with the front-left and upper-front-right regions removed for visibility purposes. The top, middle, and bottom slices are marked with **a**, **b** and **c**, respectively, corresponding to Fig. 3 **a-c**. A cube is drawn for scale in the bottom left with dimensions $20 \times 20 \times 20$ nm$^3$. Upper left: Cross-sectional slice (outlined by a dark blue rectangle) of the normalized magnetic field with the domain walls of the hybrid antiskyrmion string roughly outlined in light blue and red, and the normalized spin texture core magnetic induction defined as $B_z < -0.2$ shaded. Upper right: Area of the core spins $A_{core}$ [nm$^2$], plotted as a function of $z$ [nm].

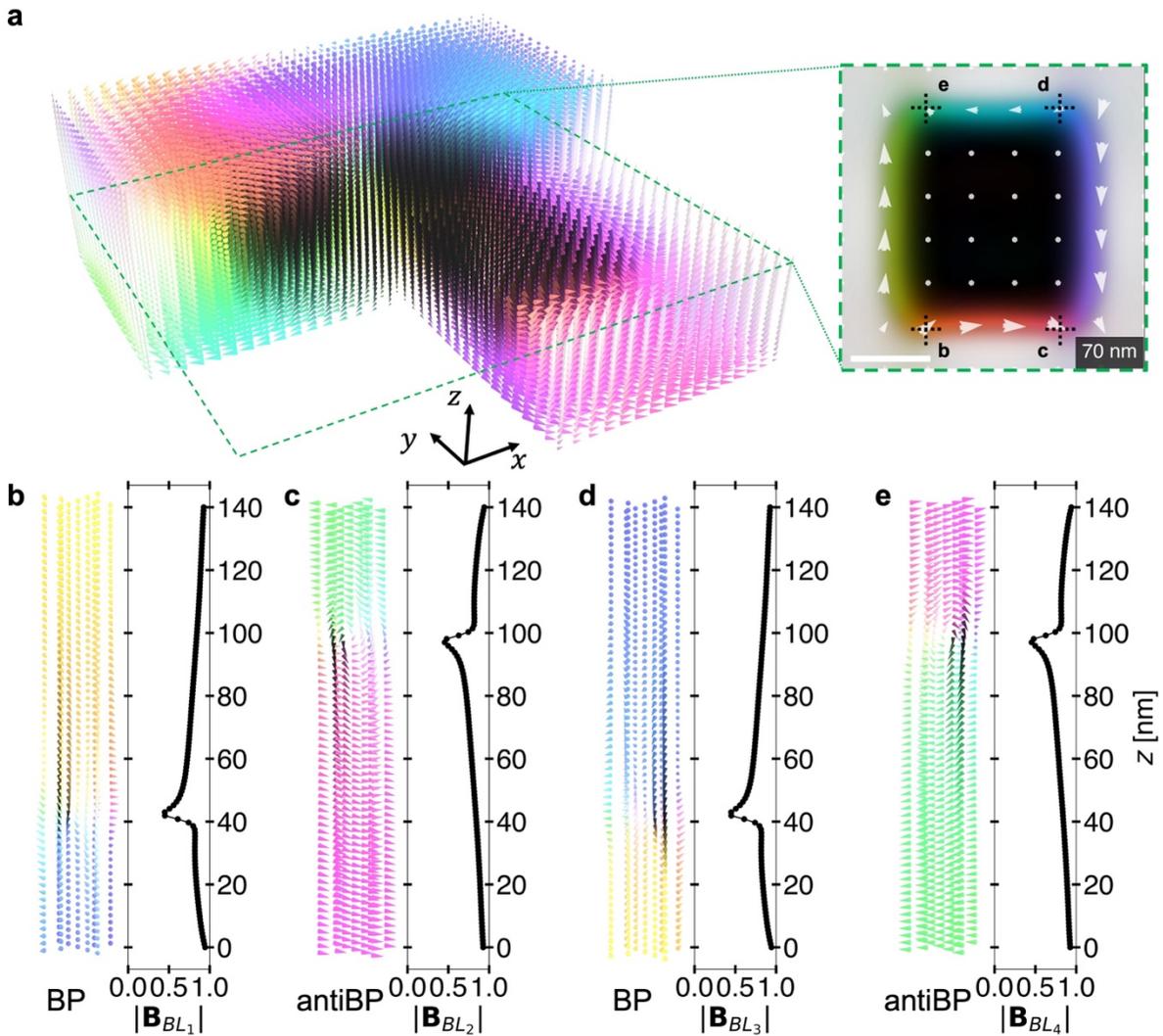

**Extended Data Figure 1. Detailed 3D structure of the simulated magnetic induction of zero field hybrid antiskyrmion strings and their BPs. a**, 3D magnetic induction $\boldsymbol{B} = \mu_0 \boldsymbol{M} + \boldsymbol{B}_{demag}$ plotted using glyphs with the front left quadrant and upper front right quadrant omitted for ease of visibility of the core structure and antiskyrmion domain walls. Right panel displays a slice through the center of the thickness, scale bar 100 nm. **b-e**, $xz$- and $yz$-slices of $\boldsymbol{B}$ intersecting the (anti)BPs positioned at the $xy$ locations indicated by dashed black crosses in the inset of **a** plotted side-by-side with the magnitude of $\boldsymbol{B}$ versus $z$ along the Bloch line which hosts the BP.

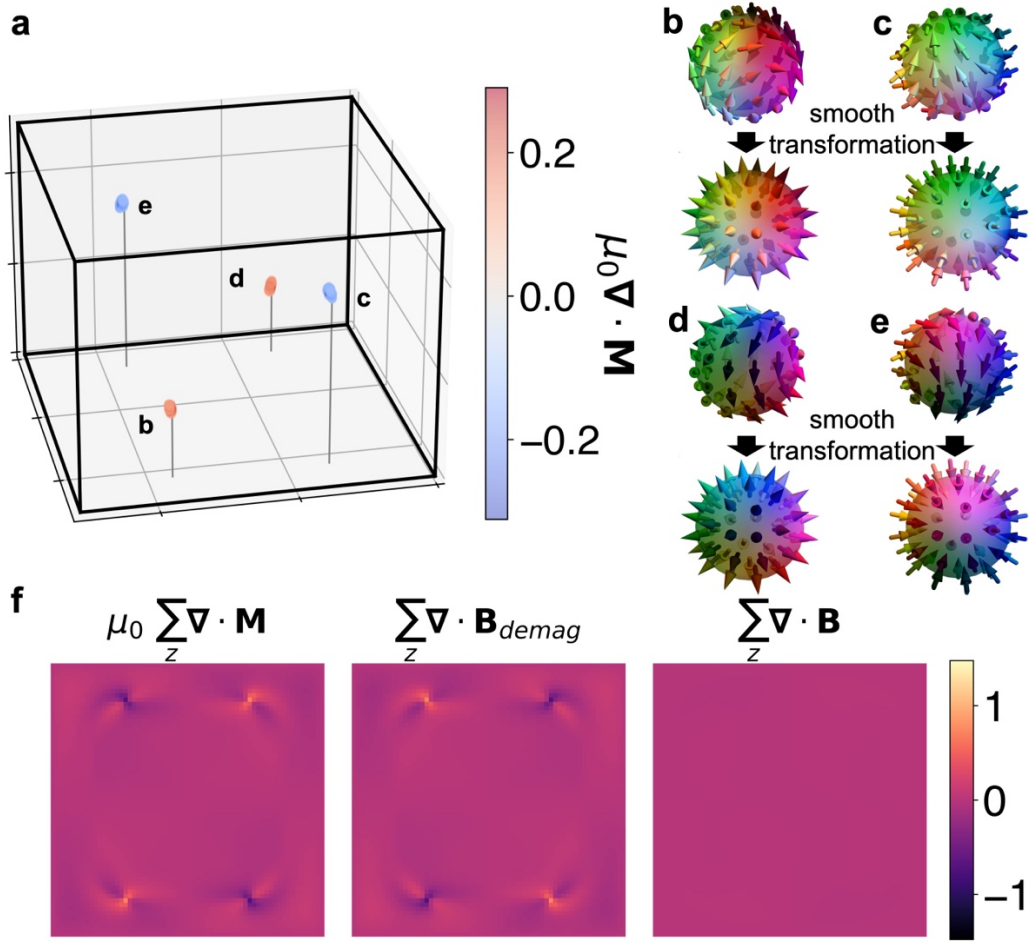

**Extended Data Figure 2. Detailed 3D structure of the simulated magnetization of the BP quadrupole. a**, 3D divergence of the magnetization, $\mu_0 \nabla \cdot M$. The data points where $\nabla \cdot M < (\nabla \cdot M)^{max}/2.5$ are omitted for ease of visibility of the (anti)BP structure. **b-e**, BPs (**b**, **d**) and antiBPs (**c**, **d**) indicated in **a** magnified with both color and glyph plots. The upper panels show the simulated magnetizations as shown in Fig. 2 (**b-e**) in the main text, while the lower panels show the same spins smoothly transformed (all spins rotated smoothly, holding the topology constant) while keeping the same color to easily track the spins, revealing the topological equivalence of these (anti)BPs with (anti)hedgehogs. The divergence in **a** also confirms this equivalence, with $\nabla \cdot M > 0$ for BPs and $\nabla \cdot M < 0$ for antiBPs, respectively. **f**, Plots of $\mu_0 \sum_z \nabla \cdot M$, $\sum_z \nabla \cdot B_{demag}$, and $\sum_z \nabla \cdot B$ from left to right, respectively, confirming the (anti)BP quadrupole constituting these hybrid spin textures.

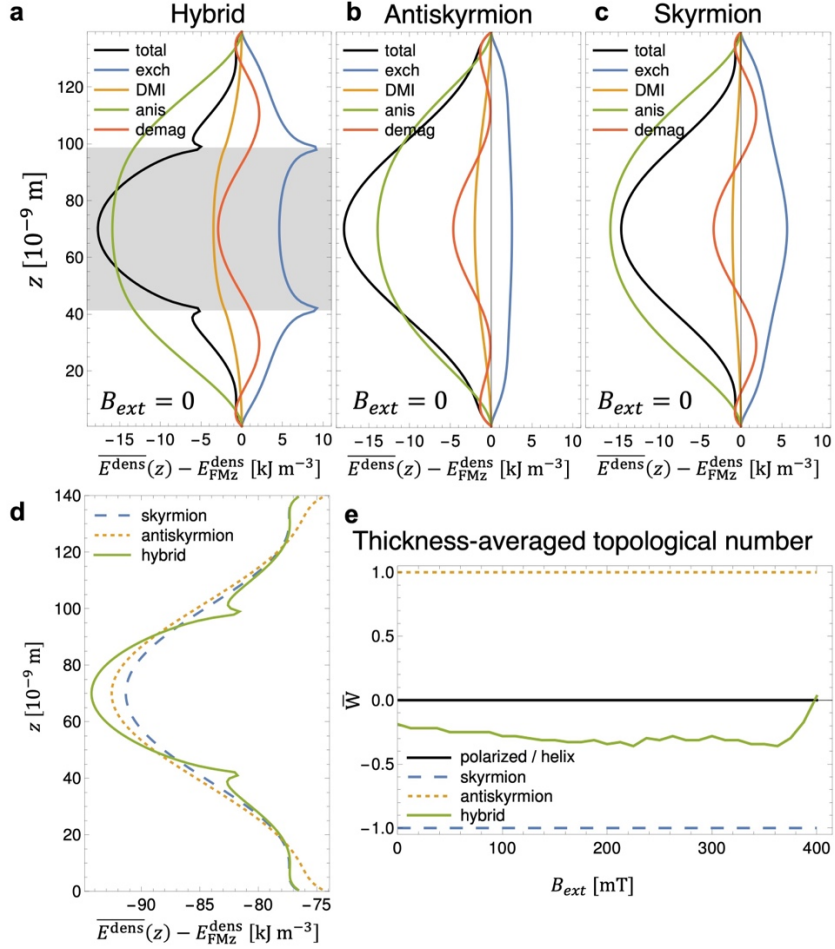

**Extended Data Figure 3. Detailed comparison between hybrid antiskyrmion strings and ideal (anti)skyrmion strings. a-c**, Total volume energy density $E_{tot}/V$, exchange energy density $E_{ex}/V$, DMI energy density $E_{dmi}/V$, uniaxial anisotropy energy density $E_{ani}/V$, and dipole-dipole energy density $E_{dd}/V$ as a function of $z$ for (**a**) hybrid antiskyrmion strings, (**b**) ideal antiskyrmion strings and (**c**) ideal skyrmion strings. The gray-shaded thickness region in **a** indicates $W = +1$, while the neighboring regions above and below indicate $W = -1$. **d**, Total volume energy density $E/V$ at $B_{ext} = 0$ as function of the penetration depth $z$ (on the y-axis). The hybrid antiskyrmion string solution gains energy (density) in the bulk like an antiskyrmion and at the surfaces like a skyrmion but has an energy penalty in the topological transition region where the BPs are located. **e**, Thickness-averaged topological number as function of magnetic field $B_{ext}$, respectively. The topology of ideal skyrmion and antiskyrmion strings is non-trivial but well-defined, +1 or -1 throughout the sample. In contrast, hybrid antiskyrmion strings have a non-integer average topological number which depends on the external magnetic field and other energy contributions.

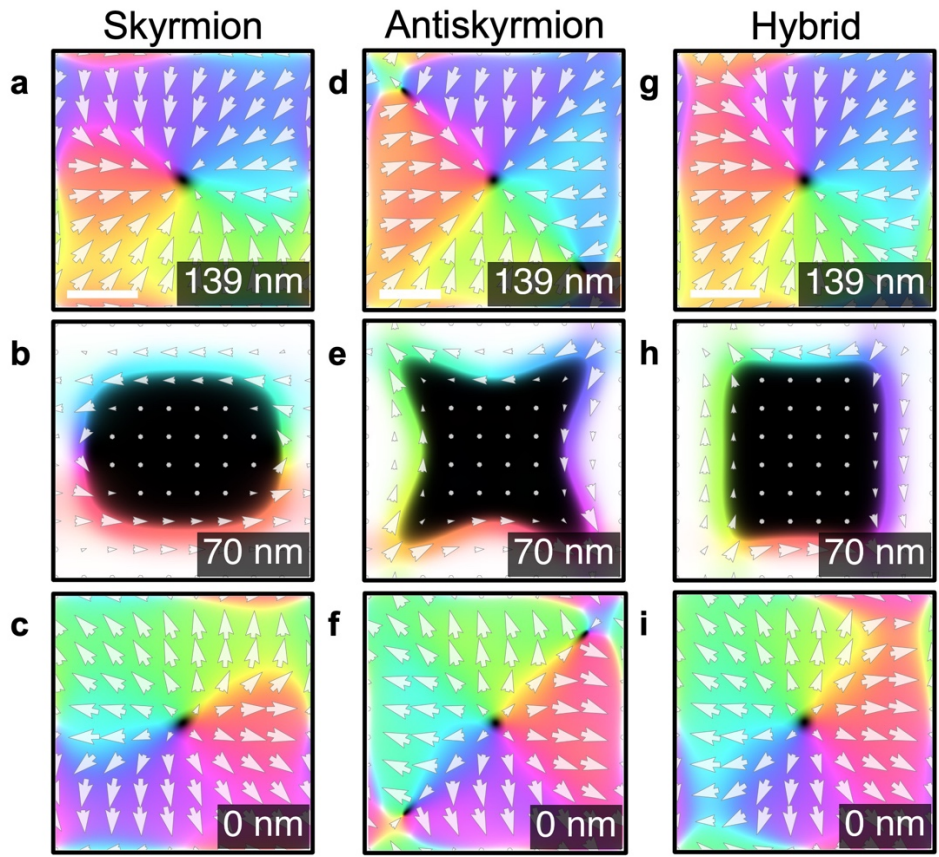

**Extended Data Figure 4: Comparison of simulated spin textures. a-i**, Selected slices of the simulated magnetic field for a (**a-c**) skyrmion string, (**d-f**) antiskyrmion string and (**g-i**) hybrid antiskyrmion string. The $xy$-slices at the (**a**, **d**, **g**) top surface, (**b**, **e**, **h**) middle, and (**c**, **d**, **i**) bottom surface of the spin textures are plotted side by side for comparison. Note the deformation at the antiskyrmion string's center (**e**) induced by the domain walls pinching along opposing Bloch lines at the antiskyrmion string's surfaces (**d**, **f**). Scalebars 100 nm.

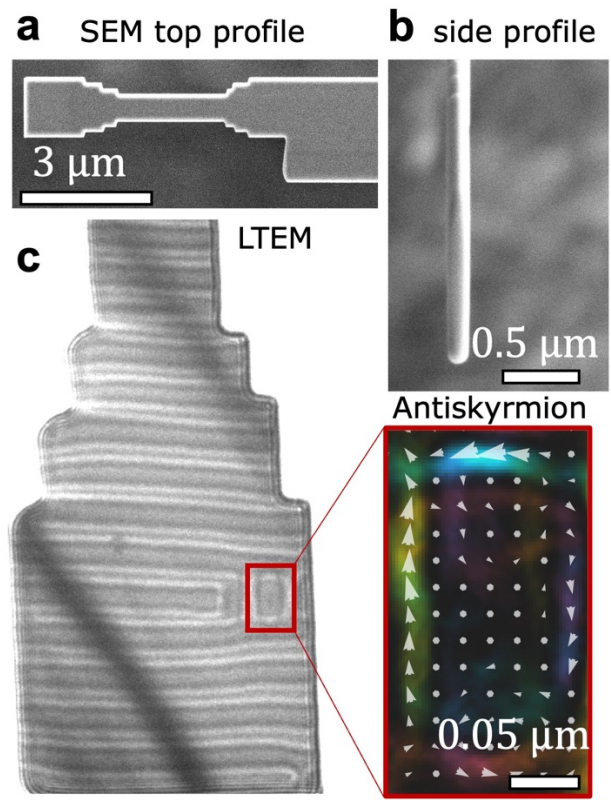

**Extended Data Figure 5: Hybrid antiskyrmion stabilized in FNPP nanoplank. a**, Scanning transmission electron microscope (SEM) micrograph of the top profile of the fabricated nanoplank. **b**, SEM micrograph of the side profile revealing a uniform thickness of ≈ 140 nm. **c**, Defocussed Lorentz TEM image of the nanoplank with the hybrid antiskyrmion indicated by a solid red rectangle. Right-hand-side panel shows the in-plane magnetic induction reconstructed via the transport of intensity equation using an over- and under-focus image.

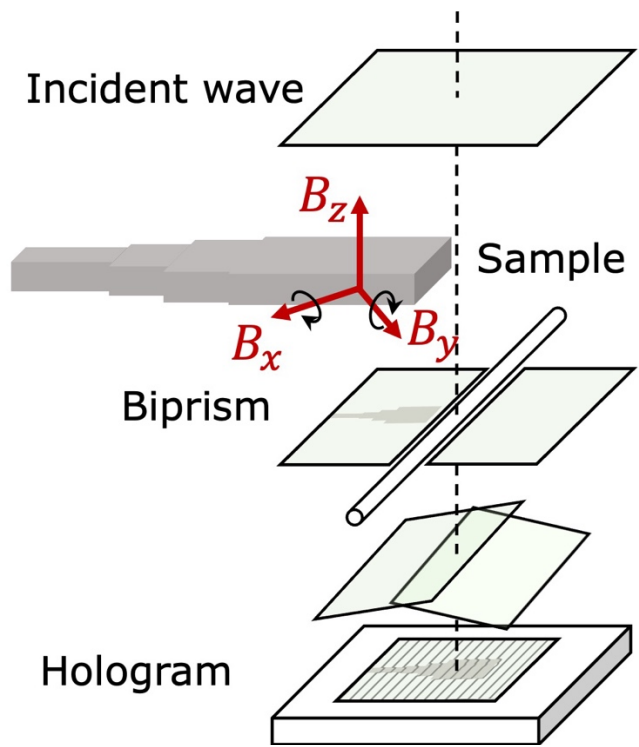

**Extended Data Figure 6: Holographic vector field electron tomography experimental setup.** The incident plane wave of electrons penetrates through the sample such that half of the beam passes through vacuum before the wavefronts are split by an electron biprism which redirects the two halves of the beam to interfere, forming an electron hologram at the direct electron detector.

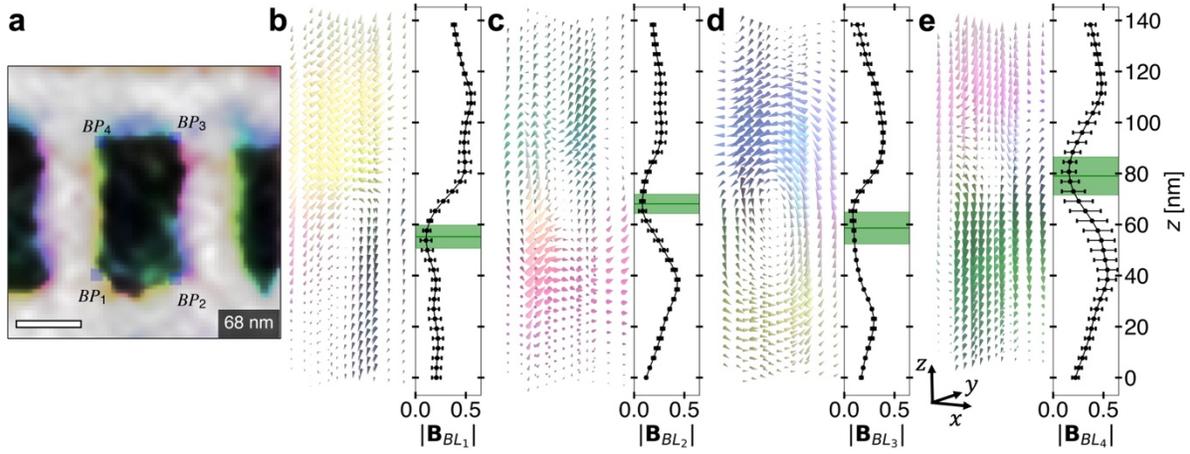

**Extended Data Figure 7: Experimentally measured 3D structure of $B(z)$ along the Bloch lines of a zero-field hybrid antiskyrmion string. a**, $xy$-slice through the center of the thickness with the approximate $xy$-location of the (anti)BPs indicated by the shaded blue box, scale bar 100 nm. **b-e**, $xz$- and $yz$-slices of $B$ intersecting the (anti)BPs positioned at the $xy$ locations indicated by the shaded blue regions in **a** plotted side-by-side with the magnitude of $B$ versus $z$ averaged over the $xy$-locations indicated in the shaded blue box in **a**, with the error bars representing the standard deviation from the mean. The solid green line represents the $z$ location of the minimum magnitude of $B$ averaged over the $xy$-locations indicated in the shaded blue box in (A), $\overline{|B|_{min}(z)}$, which may be thought of as the locations of the (anti)BPs. The $xy$-locations where $\overline{|B|_{min}(z)} > (|B|_{min}(z) + 0.1)$ were omitted from the average calculation and the shaded green region indicates the standard deviation from the mean.

| **Vector field electron tomography procedure.** |
|---|
| 1. Acquire electron holograms over 360° in 5° steps along two tilt axes. |
| 2. Reconstruct the phase maps for each tilt angle |
| 3. Assemble the series of phase maps for the a) $-65°: +65°$ and b) $+115°: -115°$ tilt angles along tilt axis x, and the c) $-63°: +67°$ and d) $+113°: -117°$ along tilt axis y. |
| 4. Align and subtract each phase map in the (b/d) series from the (a/c) series, then divide by two to obtain the magnetic phase. |
| 5. Calculate the projected magnetic induction from each phase map at all angles and retain the y-component from the x axis tilt series (a/b) and the x-component from the y axis tilt series (c/d). |
| 6. Align the magnetic induction tilt series images before using QURT to reconstruct $B_y(\mathbf{r})$ and $B_x(\mathbf{r})$ from the x axis and y axis tilt series, respectively. |
| 7. Align the $B_x$ and $B_y$ data sets manually, then calculate $B_z$ from the $\nabla \cdot \mathbf{B}(\mathbf{r}) = 0$ condition. |

**Extended Data Table 1: Vector field electron tomography procedure.**